\documentstyle[aps,preprint]{revtex}
\begin{document}
\draft
\title{Towards a theory of Warm Inflation of the Universe}
\author{Mauricio Bellini\footnote{E-mail address: ma$\_$bellini@hotmail.com}}
\address{Departamento de F\'{\i}sica, Facultad de Ciencias Exactas  y
Naturales \\ Universidad Nacional de Mar del Plata, \\
Funes 3350, (7600) Mar del Plata, Buenos Aires, Argentina.\\ 
and \\
Instituto de F\'{\i}sica y Matem\'aticas, \\
Universidad Michoacana de San Nicol\'as de Hidalgo, \\
58060 Morelia, Michoac\'an, M\'exico.}
\maketitle

\begin{abstract}
The warm inflation scenario is an alternative mechanism
which can explain the isotropic and homogeneous Universe
which we are living in.
In this work I extend a previously introduced formalism,
without the restriction of slow - roll regime.
Quantum to classical transition of the fluctuations is studied
by means of the ``transition function'' here introduced. I found
that the fluctuations of radiation energy density decrease
with time and the 
thermal equilibrium at the end of inflation holds.
\end{abstract}
\vskip 3cm
\section{Introduction}
In the old inflation scenario\cite{1}, it was assumed
that the Universe underwent isentropic expansion during the stage of
rapid growth of the scale factor. However, this scenario predicts
an inhomogeneous Universe due to the fact 
the temperature is larger than
the critical one $T_c \sim 10^{15}$ GeV.

The chaotic inflation scenario describes a quasi - de Sitter expansion
in a supercooled scenario. The entropy required to make the
post-inflationary Universe consistent with observation is assumed to be
generated in a short-time reheating period\cite{2}. However, the fluctuations
of temperature should be very large
and many domains should surpass the critical temperature $T_c$.
This fact would lead to a very inhomogeneous Universe.

Recently, Berera and Fang\cite{3} showed how thermal fluctuations
may play the dominant role in producing the initial perturbations
during inflation. They invoked slow - roll conditions. This
ingenious idea was extended in some papers\cite{4} into the warm
inflation scenario. This scenario served as an explicit demonstration that
inflation can occur in presence of a thermal component. However, the
radiation energy density $\rho_r$ must be small with respect to the
matter energy density $\rho_{\varphi}$. More exactly, the kinetic
component of the energy density ($\rho_{kin}$)
must be small with respect to the vacuum
energy density. This condition is satisfied if
\begin{equation}
\rho_{\varphi} \sim V(\varphi) \gg \rho_r \gg \rho_{kin},
\end{equation}
where $V(\varphi)$ is the potential associated with the scalar field
$\varphi$. A scenario of this kind
provides a rapid expansion of the Universe
in presence of a thermal component with small fluctuations of temperature
compatible with the COBE data\cite{5}. The thermal equilibrium is reached
near the minimum of the potential $V(\varphi)$. Particles are created
during this expansion and it is not necesary to have
a further reheating era (like in standard inflation).
More recently\cite{YL}, J. Yokoyama and A. Linde showed
that the solutions of
warm inflation violate the adiabatic condition that the scalar
field should not change significantly in the relaxation time
of particles interacting with it. They claim that if the energy
released by the interaction of the field with the created particles is
small, then the total number of particles in the warm universe must be
very small and their interaction with the scalar field may be
too small to
keep it from rapid falling down. 

According to the inflationary scenario, the inhomogeneities in the
very early Universe are of genuine quantum origin\cite{6}.
But, at end
of inflation the Universe becomes classical. The quantum to classical
transition is due to accelerated expansion of the Universe and loss of
coherence of the quantum fluctuations. This last is produced by
the increment of the degrees of freedom of the coarse - grained field which
characterizes the infrared sector, jointly with the
dissipation produced by the interaction of the inflaton with the thermal
bath\cite{8,9,10}.

In this paper, I extend the formalism of warm inflation recently
introduced\cite{9}. Exact solutions for the power - law expansion
of the Universe are calculated for this formalism. 
The paper is organized as follow: in
section II, I develope the
formalism. In section III the potential inflation model is studied,
and finally in section IV, the final comments
and conclusions are presented.

\section{The Formalism}
We consider a scalar field $\varphi$ coupled minimally to a classical
gravitational one with a Lagrangian:
\begin{equation}
{\cal L}(\varphi,\varphi_{,\mu}) = - \sqrt{-g}\left[\frac{R}{16 \pi}+
\frac{1}{2} g^{\mu \nu} \varphi_{,\mu} \varphi_{,\nu}+V(\varphi) \right]+
{\cal L}_{int},
\end{equation}
where $R$ is the scalar curvature, $g^{\mu \nu}$ the metric tensor,
$V(\varphi)$ the scalar potential and $g$ is the metric. The Lagrangian
${\cal L}_{int}$ takes into account the interaction of $\varphi$ with
other fields (i.e. bosons $X$,$Y$ or fermions) of the thermal bath. 
The spacetime will be
considered as isotropic and homogeneous, and characterized by a Friedmann -
Robertson - Walker (FRW) metric
\begin{equation}
ds^2 = - dt^2 + a^2(t) \  dr^2.
\end{equation}
The scale factor $a(t)$, is a time dependent function which grows
with time. The equation of motion
for the operator $\varphi$ is
\begin{equation}\label{1}
\ddot\varphi - \frac{1}{a^2} \nabla^2 \varphi + \left(
3 H(\varphi)+\tau(\varphi)\right) \dot\varphi+V'(\varphi) =0,
\end{equation}
where $H$ is the Hubble parameter (the dot denotes
the time derivative), $\tau(\varphi)\dot\varphi $ describes the density
energy dissipated by the field $\varphi$ into a thermalized bath,
and the prime denotes the derivative with respect to $\varphi$.

The imposibility to solve the equation (\ref{1}) becomes from
our unknowledge of the 
Hilbert's space
over which acts the field $\varphi$. For our proposal,
will be sufficient to write the Friedmann equations in the semiclassical
form
\begin{equation}
H^2 = \frac{8 \pi}{3}G \left< \rho_{\varphi}+\rho_{r}\right>,
\end{equation}
where $G=M^{-2}_p$ is the gravitational constant, $M_p$ is the Planckian
mass ($M_p = 1.2 \  10^{19}$ GeV), 
and $\rho_{r}$ and
$ \rho_{\varphi}$ are the radiation and matter energy densities
\begin{eqnarray}
&& \rho_{\varphi}= \frac{\dot\varphi^2}{2}+\frac{1}{a^2}\left(
\vec{\nabla}\varphi\right)^2+ V(\varphi), \\
&& \rho_{r} =   \frac{\tau}{8 H} \dot\varphi^2.
\end{eqnarray}
In the conventional approach to the inflaton dynamics, the field $\varphi$
is split into a spatially homogeneous classical piece plus a spatially
inhomogeneous quantum piece that represents quantum fluctuations of the
field\footnote{These fluctuations will be consider as very small.}
\begin{equation}
\varphi(\vec{x},t)= \phi_c(t) + \phi(\vec{x},t).
\end{equation}
We require that $<E|\varphi|E>=\phi_c(t)$ and $<E|\phi|E>=0$, where
$|E>$ is an arbitrary state.

\subsection{Dynamics of the classical field $\phi_c$}

We define the classical field as a solution of the equation
\begin{equation}
\ddot\phi_c+ (3 H_c + \tau_c) \dot\phi_c+ V'(\phi_c)=0.
\end{equation}
The Hubble parameter is expanded as:
\begin{equation}
H= H_c \left[ 1 + \frac{1}{2 H^2_c} \left< \left(1+
\frac{\tau_c}{4 H_c}\right)
\dot\phi^2+ \frac{1}{a^2}(\vec{\nabla}\phi)^2
+ \sum_n \frac{1}{n!} V^{(n+1)}
(\phi_c) \  \phi^n \right> \right],
\end{equation}
where the classical Hubble parameter is
\begin{equation}
H^2_c = \frac{4 \pi}{3 M^2_p} \left[\left( 1+ \frac{\tau_c}{4 H_c}\right)
\dot\phi^2_c+ 2 V(\phi_c)\right].
\end{equation}
The classical dynamics of $\phi_c$ and $H_c$ are characterized by the
following equations
\begin{eqnarray}
\dot\phi_c &=& - \frac{M^2_p}{4\pi} H'_c \left( 1
+ \frac{\tau_c}{3 H_c}\right)^{-1}, \\
\dot H_c &=& H'_c \  \dot\phi_c =   - \frac{M^2_p}{4\pi} (H'_c)^2
\left( 1+ \frac{\tau_c}{3 H_c}\right)^{-1},
\end{eqnarray}
and the potential is
\begin{equation}
V(\phi_c)= \frac{3 M^2_p}{8\pi} \left[ H^2_c - \frac{M^2_p}{12\pi}
(H'_c)^2 \left( 1+ \frac{\tau_c}{4 H_c}\right)
\left( 1+ \frac{\tau_c}{3 H_c}\right)^{-2} \right],
\end{equation}
where we have assumed $H(\varphi)=H(\phi_c)\equiv H_c$ and
$\tau(\varphi)=\tau(\phi_c)\equiv \tau_c$.
The expression for radiation energy density is
\begin{equation}\label{ro}
\rho_r \simeq \frac{\tau_c}{8 H_c} \left(\frac{M^2_p}{4\pi}\right)^2
(H'_c)^2 \left( 1+ \frac{\tau_c}{3 H_c}\right)^{-2}.
\end{equation}
Near the minimum of the potential, the thermal equilibrium holds
and the temperature is
\begin{equation}
< T_r > \sim \left( \frac{\tau_c(0)}{4 H(\phi_c=0)}
\dot\phi^2_c\right)^{1/4}.
\end{equation}
In the inflationary epoch the kinetic energy 
is much smaller than the vacuum energy density $\rho_{\varphi}(\phi_c) \sim
V(\phi_c)$.
As in a previous work\cite{9}, I will consider the
following relation between the Hubble parameter $H_c(\phi_c)=\dot a/a$ and
the friction one $\tau_c(\phi_c)$:
\begin{equation}\label{ga}
\tau_c(\phi_c) = \gamma \  H_c(\phi_c).
\end{equation}
This expression takes into account that the particlelike are
dispersed during the expansion of the Universe. The rate
of expansion of the Universe is given by the Hubble parameter
$H_c(\phi_c)$ and the increment of the rate of expansion
lead to an increment of the rate of friction. The dimensionless
constant $\gamma = \tau_c / H_c$ in the equation 
(\ref{ga}) is a parameter of the 
theory which takes into account the intensity of friction due to
the interaction between the inflaton and the fields of the thermal
bath. Note that in a the de Sitter expansion $H_c(\phi_c)$
is constant (i.e., $H_c(\phi_c)=H_o$), and thus $\tau_c$ and $\rho_r$
become zero ($\tau_c=\rho_r=0$). In this case one recovers the standard
inflation model\cite{10}. 
The dynamics for the expansion and friction parameters is governed by the
evolution of the classical field $\phi_c$.

\subsection{Dynamics of the Quantum Fluctuations}

The equation of motion for the operator $\phi$ is
\begin{equation}
\ddot\phi - \frac{1}{a^2} \nabla^2 \phi + (3 H_c+\tau_c)\dot\phi+
\sum_n \frac{1}{n!} V^{(n+1)}(\phi_c) \phi^n =0,
\end{equation}
where $V^{(n+1)}(\phi_c)$ denotes the $n+1$-th derivative. We assume
that the quantum fluctuations are small and the Taylor
expansion of $V'(\varphi)$ can be truncated at first order
in $\phi$
\begin{equation}
\ddot\phi - \frac{1}{a^2} \nabla^2 \phi + (3 H_c+\tau_c)\dot\phi+
V''(\phi_c) \phi =0.
\end{equation}

For our propose, will be useful to consider the redefined field
$\chi = e^{3/2 \int (H_c+\tau_c/3) \  dt} \  \phi$. The 
equation of motion for this field is
\begin{equation}
\ddot\chi- a^{-2} \nabla^2 \chi - \frac{k^2_o}{a^2} \chi =0,
\end{equation}
where
\begin{equation}
k^2_o= a^2 \left[ \frac{9}{4} \left( H_c+\frac{\tau_c}{3}\right)^2 -
V''(\phi_c) + \frac{3}{2}\left( \dot H_c
+ \frac{\dot\tau_c}{3}\right) \right],
\end{equation}
is the squared time dependent wavenumber. The field $\chi$ can be written as
a Fourier expansion of the modes $\xi_k(t) \  e^{i \vec{k}.\vec{r}}$
\begin{equation}
\chi = \frac{1}{(2\pi)^{3/2}} \int d^3 k \left[ a_k e^{i \vec{k}.\vec{r}}
\xi_k(t) + h.c. \right],
\end{equation}
where $\xi_k(t)$ are the time dependent modes. 
The annihilation and creation operators $a_k$ and $a^{\dagger}_k$ satisfy
the commutation relations
\begin{eqnarray}
&& [a_k,a^{\dagger}_{k'}]  =  \delta (\vec{k} - \vec{k'}), \\
&& [ a_k, a_{k'} ]  =  [a^{\dagger}_k,a^{\dagger}_{k'}] = 0.
\end{eqnarray}
The equation of motion for $\xi_k$ is
\begin{equation}
\ddot\xi_k(t)+ \omega^2_k \  \xi_k(t) =0,
\end{equation}
where $\omega^2_k = a^{-2}(k^2-k^2_o)$ is the squared frecuency
of each mode with a given wavenumber $k$.
The function $k_o$ separates both, the unstable $(k^2 \ll k^2_o)$ 
and the stable $(k^2 \gg k^2_o)$ sectors. The quantum 
fluctuations with wavenumbers
below $k_o$ are interpreted as inhomogeneities superimposed on the
classical field $\phi_c$. These fluctuations are
responsible for the density inhomogeneities
generated during the inflation. The quantum field theory imposes the
commutation relation between $\chi$ and $\dot\chi$
\begin{equation}\label{2}
[\chi(\vec{r},t),\dot\chi(\vec{r'},t)] = {\rm i} 
\delta^{(3)} \  (\vec{r}-\vec{r'}).
\end{equation}
The equation (\ref{2}) implies that
\begin{equation}
\xi_k \dot\xi^*_k - \dot\xi_k \xi^*_k = {\rm i}.
\end{equation}
To study the Universe on a scale greater
than the scale of the observable Universe, we define the coarse - grained 
field that takes into
account the long - wavelength modes. The wavelengths of such
modes are
\begin{equation}
l \ge \frac{1}{\epsilon k_o},
\end{equation}
where $\epsilon \ll 1$ is a dimensionless constant. 
The coarse - grained field, so defined, is
\begin{equation}
\chi_{cg}(\vec{r},t) = \frac{1}{(2\pi)^{3/2}} \int d^3 k \  \theta
(\epsilon k_o -k)
\left[ a_k e^{i\vec{k}.\vec{r}} \xi_k(t) + h.c. \right],
\end{equation}
which satisfies the following operatorial stochastic equation
\begin{equation}\label{p1}
\ddot \chi_{cg} - \frac{k^2_o}{a^2} \chi_{cg} = \epsilon
\left(\frac{d}{dt}(\dot k_o \eta) + 2 \dot k_o \kappa\right),
\end{equation}
with the operatorial noises
\begin{eqnarray}
\eta(\vec{r},t) &=& \frac{1}{(2\pi)^{3/2}} 
\int d^3k \  \delta(\epsilon k_o - k) 
\quad \left[a_k e^{i\vec{k}.\vec{r}} \xi_k(t) + h.c \right], \\
\kappa(\vec{r},t) &=& \frac{1}{(2\pi)^{3/2}} 
\int d^3k \  \delta(\epsilon k_o - k)   
\quad\left[a_k e^{i\vec{k}.\vec{r}} \dot\xi_k(t) + h.c \right].
\end{eqnarray}
The commutation relations are
\begin{eqnarray}
&& [\chi_{cg}(t), \eta(t)] = 0, \label{c2}\\
&& [\chi_{cg}(t), \kappa(t)]  =
\frac{2 \epsilon k_o}{(2\pi)^{3}} \int d^3 k \quad
\theta(\epsilon k_o - k) \delta(\epsilon k_o - k) \quad
\left( \xi_k \dot\xi^*_k - \xi^*_k \dot\xi_k\right), \label{c3}\\
&& [\eta(t), \kappa(t)] =
\frac{2 \epsilon k_o}{(2\pi)^{3}} \int d^3 k \quad
\delta(\epsilon k_o - k) \delta(\epsilon k_o - k) 
\quad \left( \xi_k \dot\xi^*_k - \xi^*_k\dot\xi_k
\right).\label{c4}
\end{eqnarray}
The correlations between the noises are
\begin{eqnarray}
&&<\kappa(t) \kappa(t') > =\frac{1}{(2\pi)^{3}}
\int d^3 k \quad \delta(\epsilon k_o(t) - k)
\quad \delta(\epsilon k_o(t') - k) 
\quad \left( \dot\xi_k(t) \dot\xi^*_k(t') \right),\label{a} \\
&&<\eta(t) \eta(t') > =\frac{1}{(2\pi)^{3}}
\int d^3 k \quad \delta(\epsilon k_o(t) - k)
\quad \delta(\epsilon k_o(t') - k) 
\quad\left( \xi_k(t) \xi^*_k(t') \right), \label{b} \\
&& <\eta(t)\kappa(t')+\kappa(t) \eta(t') >
 = \frac{2}{(2\pi)^{3}} \int d^3 k \quad
\delta(\epsilon k_o(t) - k) \quad
\delta(\epsilon k_o(t') - k) 
\quad \left( \dot\xi_k(t) \xi^*_k(t') \right).\label{c}
\end{eqnarray}
Quantum to classical transition occurs due to the rapid expansion of the
Universe during inflation. This
transition is satisfied when the commutators (\ref{2}), (\ref{c2}),
(\ref{c3}) and (\ref{c4})
are nearly zero. We can write the time dependent modes as
\begin{equation}
\xi_k(t)= u_k(t)+ {\rm i} \  v_k(t),
\end{equation}
and the condition to obtain the complex to real transition of these modes is
\begin{equation}
\left|\frac{v_k(t)}{u_k(t)}\right| \ll 1.
\end{equation}
We can define the quantum to classical transition function
$\alpha_k(t)=\left|{v_k(t)\over u_k(t)}\right|$.
The modes are real when this function becomes nearly zero.
The condition for the coarse - grained field $\chi_{cg}(\vec{r},t)$
to be classical is
\begin{equation}
\frac{1}{N(t)}\sum_{k=0}^{k= \epsilon k_o}  \alpha_k(t) \ll 1,
\end{equation}
where $N(t)$ is time dependent
number of degrees of freedom of the infrared sector.
During inflation, this number increases with time.
This effect is due to the
temporal evolution of both, the scale factor $a$ and
the superhorizon (with size $l \sim
{1 \over \epsilon k_o}$).
Quantum to classical
transition of the coarse-grained field holds when
$\alpha_{k=\epsilon k_o}(t)\rightarrow 0$.

Making ${\rm tan}[\Theta_k(t)]={v_k(t)\over u_k(t)}$, we
can write the modes as
\begin{equation}
\xi_k(t)= e^{i \Theta_k(t)} \quad \left|\xi_k(t)\right|,
\end{equation}
where $\left|\xi_k(t)\right|= \sqrt{v^2_k+u^2_k}$.
When the quantum fluctuations become classical, the correlations
(\ref{a}), (\ref{b}) and (\ref{c}) are
\begin{eqnarray}
&&<\kappa(t) \kappa(t') > =\frac{1}{2\pi^{2}}
\frac{\epsilon k^2_o}{\dot k_o}
\xi_k(t)\xi_k(t') \quad\delta(t - t') \\
&&<\eta(t) \eta(t') > = \frac{1}{2\pi^{2}}
\frac{\epsilon k^2_o}{\dot k_o}
\dot\xi_k(t)\dot\xi_k(t') \quad\delta(t - t') \\
&& <\eta(t)\kappa(t')+\kappa(t) \eta(t') >
= \frac{1}{2\pi^{2}}
\frac{\epsilon k^2_o}{\dot k_o}
\left(\xi_k(t)\dot\xi_k(t')+\dot\xi_k(t)\xi_k(t')\right)\quad
\delta(t - t').
\end{eqnarray}

\subsection{Coarse-grained field correlations and power
spectral density}

Now we consider the correlation of $\chi_{cg}$ for different times $t$
and $t'$ ($t'> t$), once the modes are classical [$\xi_k(t)\xi^*_k(t')
\simeq \xi_k(t)\xi_k(t')$]
\begin{equation}
< \chi_{cg}(t) \chi_{cg}(t')> = \frac{1}{(2\pi)^3}
\int^{\epsilon k_o(t')}_{\epsilon k_o(t)} d^3  k \quad \xi_k(t)\xi_k(t').
\end{equation}
The Fourier transform of $< \chi_{cg}(t) \chi_{cg}(t')>$ gives
the power spectral density for $\chi_{cg}$ in the infrared sector
\begin{equation}
S[\chi_{cg};\omega_k]= 4 \int^{\infty}_{0}
cos[\omega t''] \quad \left|< \chi_{cg}(t) \chi_{cg}(t+t'')>\right|_{t\gg 1}
\quad dt'',
\end{equation}
where $t''=t'-t$. This expression, written explicitly, becomes
\begin{equation}\label{S}
S[\chi_{cg};\omega_k]=\frac{1}{\pi}\int^{\infty}_{0} dt''\quad
cos[\omega t''] \quad \int^{\epsilon k_o(t+t'')}_{\epsilon k_o(t)}
dk \  k^2 \  \xi_k(t)\xi_k(t+t''),
\end{equation}
where $\omega_k$ is the oscillation frecuency for a given wavenumber $k$.
The equation (\ref{S}) is only valid when the thermal equilibrium holds.
The fluctuations of 
radiation energy density are
\begin{equation}
\frac{\delta\rho_r}{\rho_r} \simeq \left| 
2(H'_c)^{-1} \  H''_c \right| \quad <\phi^2_{cg}>^{1/2},
\end{equation}
where $\phi_{cg}(\vec{r},t)=a^{-1/2(3+\gamma)} \  \chi_{cg}(\vec{r},t)$.
The condition for the thermal equilibrium holds is
${d\over dt}\left({\delta\rho_r \over \rho_r}\right)< 0$.

\section{Power - law Inflation}

This model of inflation is characterized by $a(t)$ and $H_c$ given by
\begin{eqnarray}
&& a(t)= H^{-1}_o \  \left(\frac{t}{t_o}\right)^p, \\
&& H_c(t)= \frac{p}{t}.
\end{eqnarray}
The temporal evolution of the classical field is
\begin{equation}
\phi_c(t)= \phi_o- m {\rm ln} \  \left[\frac{H_o}{p}t\right].
\end{equation}

The solution of the equation (\ref{ro}) for the radiation energy density
is
\begin{equation}
\rho_r = \frac{\gamma M^4_p \  p^2}{128 \pi^2 m^2 (1+\gamma/3)^2} \  t^{-2}.
\end{equation}
The radiation temperature is
\begin{equation}
<T_r> \propto M_p m^{-1/2} \  t^{-1/2},
\end{equation}
where $m$ is the mass of the scalar field.
The potential for
this model is
\begin{eqnarray}
&& V(\phi_c)= \frac{3 M^2_p}{8 \pi} H^2_c \left\{ 1-
\frac{M^2_p}{48 \pi} m^{-2} \left[\left( 1+\gamma /4 \right)\left(
1+\gamma/3\right)^{-2} \right] \right\}.
\end{eqnarray}
The equation of motion
for the modes is
\begin{equation}\label{mo}
\ddot \xi_k(t)
+ \left[\frac{H^2_o t^{2p}_o k^2}{t^{2p}}-\mu^2(t)\right] \quad
\xi_k(t)=0,
\end{equation}
with
\begin{equation}
\mu^2(t)=t^{-2} K^2,
\end{equation}
and
\begin{equation}\label{k}
K^2=\left[ \frac{9}{4} p^2(1+\gamma/3)^2-
\frac{3}{2}p(1+\gamma/3)+\frac{3 p^2M^2_p}{2\pi m^2}
\left(1- \frac{M^2_p}{48 \pi^2 m^2} (1+\gamma/4)(1+\gamma/3)^{-2}\right)
\right].
\end{equation}
For inflation takes place, we must take $k^2_o>0$,
and the expansion must be
sufficiently rapid such that $p> {3 \over 2 A} (1+\gamma/3)$, with
$A=9/4(1+\gamma/3)^2+ {3 M^2_p \over 2 \pi m^2} \left( 1 -
{M^2_p \over 48 \pi m^2} (1+\gamma/4)(1+\gamma/3)^{-2}\right)$.
The general solution for the equation (\ref{mo}) is
\begin{equation}
\xi_k(t)= A_1 \sqrt{t} \  H^{(1)}_{\nu}(t)\left[\frac{H_o
k (t/t_o)^{1-p}}{p-1}\right]
+ A_2 \sqrt{t} \  H^{(2)}_{\nu}(t)\left[\frac{H_o
k (t/t_o)^{1-p}}{p-1}\right],
\end{equation}
with 
$\nu = {1\over 2(p-1)} \sqrt{1+4 K^2}$. When $\gamma=0$ (i.e., for
$\tau_c=0$), we recover the results of standard inflation\cite{10}.
We choose the Bunch - Davis vacuum ($A_1=0$), but
with imaginary constant $A_2={\rm i} |A_2|$ that can be written
in terms of the Bessel functions
\begin{equation}
\xi_k(t)= \frac{\sqrt{\pi}}{2}\frac{\sqrt{t/t_o}}{p-1}
\left[{\cal Y}_{\nu}[x(t)]+{\rm i} {\cal J}_{\nu}[x(t)]\right]
={\rm i} \  H^{(2)}_{\nu}[x(t)],
\end{equation}
where $H^{(2)}_{\nu}[x(t)]$ is the Hankel function and 
$x(t)= {H_o k \left(t/t_o\right)^{1-p}\over p-1}$. Observe that $x(t)$ tends
to zero when $t$ is sufficiently large.
In this case one obtains $x(t)\ll 1$, and the asymptotic modes are real 
\begin{equation}\label{xi}
\xi_k(t) \simeq \frac{\sqrt{\pi}}{2} \frac{\sqrt{t/t_o}}{p-1}
{\cal Y_{\nu}}\left[ x(t)\right],
\end{equation}
which is due to
\begin{equation}
\left|\frac{{\cal J}_{\nu}[x(t)]}{{\cal Y}_{\nu}[x(t)]}\right| \ll 1.
\end{equation}
So, for $\left(t/t_o\right)^{p-1}\gg \frac{H_o k}{p-1}$ 
the modes with the
wavenumber $k$ are real. 
For
$x(t)\ll 1$ the function ${\cal Y}_{\nu}$ becomes
\begin{equation}
{\cal Y}_{\nu}  \simeq
- \frac{{\rm i}}{\pi} \Gamma(\nu) 
\left( \frac{H_o k (t/t_o)^{1-p}}{2 (p-1)}\right)^{-\nu},
\end{equation}
where $\Gamma(\nu)$ is the gamma function with argument $\nu$.
The correlation of the coarse - grained field, for $t\gg 1$, is
\begin{equation}\label{corr}
\left.<\phi_{cg}(t=t_o) \phi_{cg}(t_1)> \right|_{p\gg 1} \simeq
\frac{2\pi 4^{\nu} \Gamma^2(\nu)}{(p-1)^4} \left[
\epsilon K \right]^{2(1-\nu)+1} \  \left(\frac{t_1}{t_o}\right)^{4\nu
p(1-\nu/p-p/\nu)+3-\gamma},
\end{equation}
where $K$ is given by the equation (\ref{k}).

The power spectral density of the fluctuations (when the thermal
equilibrium holds) is the Fourier transform of the equation (\ref{corr}),
which becomes [see eqs. (\ref{S}) and (\ref{xi})]
\begin{eqnarray}
S[\chi_{cg}; \omega_k=\frac{k H_o }{(t_1/t_o)^p}] & \simeq & 
\frac{8\pi 4^{\nu} \Gamma^2(\nu)}{(p-1)^4} \left[
\epsilon K \right]^{2(1-\nu)+1}
\cos\left[\frac{1}{2}\pi\left(4\nu(p-1)-2p+4\right)\right]\nonumber \\
&\times & \Gamma\left[\left(4\nu(p-1)-2p+4\right)\right]
\left|\omega_k\right|^{-\left( 4\nu (p-1)-2p+4 \right) }.
\end{eqnarray}
For $p \gg 1$ one obtains $S[\chi_{cg}, \omega_k] \propto
\left|\omega_k\right|^n$, with $n= -(4\nu(p-1)-2p+4)$.
The fluctuations of the radiation energy density are
\begin{equation}
\left.\frac{\delta\rho_r}{\rho_r}\right|_{t\gg 1}
\sim t^{2\nu p(1-\nu/p-\frac{5}{4}p/\nu)
+\frac{1}{2}-\frac{\gamma}{2}}.
\end{equation}
Note that both, $<\phi_{cg}(t_o) \phi_{cg}(t_1)> $
and  ${\delta\rho_r\over\rho_r}$ decreases with
time for sufficiently large $p$.

\section{Conclusions}

In this work, I extended the formalism for warm inflation
introduced in a previous paper. In the model the
rapid expansion of the Universe is produced with a 
temperature $T_r$ smaller than $T_c$. 
The dynamics of the fluctuations of the matter field
generates
thermal fluctuations which decrease with time. So, at the
end of inflation, the thermal equilibrium holds. On 
the other hand, the classical field $\phi_c(t)$ is 
responsible for the expansion of the Universe
and the dissipation of energy. The dissipation of energy
is due to the interaction between the inflaton
and the particles in the thermal bath (with mean temperature
$<T_r> \  < T_c$). The dynamics of the interaction
is represented by the classical parameter $\tau_c(\phi_c)$.
This parameter is considered as proportional to the rate of expansion
of the Universe, which is given by the Hubble parameter $H_c(\phi_c)$.

This model has four foundamental aspects:
{\bf 1)} The classical one: characterized by the size of the
superhorizon $l(t)\sim {1\over \epsilon k_o}$.
The quantum to classical transition of the fluctuations in the infrared
sector is due
to the complex to real transition of the modes $\xi_k$ 
of $\chi_{cg}$.
The dynamics of this transition is described by the sum over
$k$ in the infrared sector, for the function $\alpha_k(t)$.
{\bf 2)} The stochastic ingredient of the coarse - grained
field: the fluctuations of the matter field $\chi_{cg}$ describes the
infrared sector and and its evolution is governed by a
classical stochastic equation.
{\bf 3)} The thermal ingredient: the dissipation parameter
$\tau_c$ represents the interaction of the inflaton with a thermal bath.
In this model, when the inflation starts the Universe is with a mean
temperature smaller than the GUT's temperature ($T_c \sim 10^{15}$ GeV).
In the example for power - law inflation here studied, the mean
temperature decreases as $t^{-1/2}$ and the thermal fluctuations
also decrease for sufficiently large $p$. So, at the end 
of inflation the Universe is more cold than when inflation starts.
The thermal equilibrium holds due to
the decreasing with time
of the fluctuations for the radiation energy density. Due to
this fact, it is possible to calculate 
the power spectral density for $\chi_{cg}$. This
spectrum depends on the rate of expansion of the Universe
and the interaction of the inflaton with
the thermal bath.
{\bf 4)} The quantum aspect: relies in
the time dependent modes $\xi_k(t)$ of the coarse - grained field
$\chi_{cg}$. These modes must be real 
in the infrared sector to give classical fluctuations.

Initially (i.e., when
the inflation starts), the 
fluctuations $\phi(\vec x, t)$ [and also $\chi(\vec x,t)$]
are quantized,
since the energy density of the false vacuum is of the order (but less)
of the Planckian
scale ($\rho_{\varphi} \sim V(\varphi) < M^4_p$).
Furthermore the amplification
of the modes that enters in the infrared sector generates
loss of coherence of the field $\chi_{cg}$.
More rapid is the
expansion of the Universe, more
rapid is the complex to real transition of each mode
with a wavenumber $k$. The modes $\xi_k$ with smaller wavenumber will
become real before the bigger ones. 
So, the Universe becomes classical in the infrared sector.
\vskip 1cm                                  
\centerline{\bf Acknowlekgements}
\vskip 1cm
I thank Adnan Bashir for your careful reading of
the manuscript.

\end{document}